\documentstyle[mprocl]{article}

\def\be{\begin{equation}}
\def\ee{\end{equation}}
\def\bea{\begin{eqnarray}}
\def\eea{\end{eqnarray}}

\begin{document}

\title{MICROLENSING IMPLICATIONS FOR HALO DARK MATTER}

\author{FRANCESCO DE PAOLIS\footnote{Also at the Bartol Research
Institute, Univ. of Delaware, Newark,Delaware, 19716-4793, USA.} 
and PHILIPPE JETZER}
\address{Paul Scherrer Institute, Laboratory for Astrophysics, CH-5232 
Villigen PSI, and
Inst. of Theor. Phys., Univ. of Z\"urich, Winterthurerstr.
190, CH-8057 Z\"urich, Switzerland}

\author{GABRIELE INGROSSO}
\address{Dipartimento di Fisica and INFN, Universit\'a di Lecce,
CP 193, I-73100 Lecce, Italy}

\author{MARCO RONCADELLI}
\address{INFN, Sezione di Pavia, Via Bassi 6, I-27100 Pavia, Italy}

\maketitle
\abstracts{
The observations of microlensing events in the Large Magellanic Cloud
suggest that a sizable fraction ($\sim$ 50\%)
of the galactic halo is in the form of
MACHOs (Massive Astrophysical Compact Halo Objects) with an 
average mass
$\sim 0.27 M_{\odot}$, assuming a standard spherical
halo model. We describe 
a scenario in which dark clusters of MACHOs and 
cold molecular clouds (mainly of $H_2$) naturally form in the halo at 
galactocentric distances larger than 10--20 kpc. }

\section{Introduction}
A central problem in astrophysics concerns the nature of the
dark matter in galactic halos, whose presence is implied by the flat 
rotation curves in spiral galaxies. As first proposed by 
Paczy\'nski \cite{pa}, 
gravitational microlensing can provide a decisive answer to 
that question \cite{kn:Derujula1},
and since 1993 this dream has started to become a reality 
with the detection of several microlensing events towards the Large 
Magellanic Cloud \cite{al,au}. 
Today, although the evidence for 
MACHOs is firm, the 
implications of this discovery crucially depend on the assumed galactic 
model. It has become customary to take the standard spherical halo model as 
a baseline for comparison. 
Within this model,
the average mass reported by the MACHO team is
$0.5^{+0.3}_{-0.2}~M_{\odot}$, which is based upon their first two years data
\cite{al}. 
The inferred optical depth is $\tau = 2.1^{+1.1}_{-0.7} \times 10^{-7}$
when considering 6 events \footnote{In fact, the two disregarded events are 
a binary lensing and one which is rated as marginal.}
(or $\tau = 2.9^{+1.4}_{-0.9} \times 10^{-7}$ when
considering all the 8 detected events). Correspondingly, this implies
that about 45\% (50\% respectively) of the halo dark matter is in form of 
MACHOs assuming a standard spherical halo model.

Instead, using the mass moment method yields 
an average MACHO mass \cite{je} of $0.27~M_{\odot}$. 
Unfortunately, because of the presently available limited statistics 
different data-analysis procedures
lead to results which are only marginally consistent. 
Apart from the low-statistics problem  --
which will automatically disappear from future larger data samples -- we 
feel that the real question is whether the 
standard spherical halo model correctly describes our galaxy
\cite{kn:Ingrosso}.
Besides the observational evidence that spiral galaxies generally have 
flattened halos, recent determinations of the disk scale length, the 
magnitude and slope of the rotation at the solar position indicate that
our galaxy is best described by the maximal disk model, which implies 
a minimal halo model.
This conclusion is further strengthened 
by the microlensing results towards the galactic centre, which 
imply that the bulge is more massive than previously thought.
For such halo models 
the expected average MACHO mass should be smaller than within the
standard halo model. Indeed, a value $\sim 0.1~M_{\odot}$ looks 
as the most realistic estimate to date and suggests that MACHOs 
are brown dwarfs.

\section{Mass moment method}
 
The most appropriate way to compute the average mass and other
important properties of MACHOs is to use
the method of mass moments developed by De R\'ujula et al. \cite{kn:Derujula}.
The mass moments $<\mu^m>$ are 
related to $<\tau^n>=\sum_{events} \tau^n$,
with $\tau \equiv (v_H/r_E) T$, as constructed
from the observations 
($v_H = 210~{\rm km}~{\rm s}^{-1}$,
$r_E = 3.17 \times 10^9~{\rm km}$ and $T$ is the duration of an event
in days). We consider only 6 (see footnote $b$)
out of the 8 events observed by the MACHO group during their first two years
\footnote{In the meantime the MACHO group has found at least six additional
events towards the LMC and at least one towards the SMC \cite{kn:SMC}.
These data are, however, not yet fully analyzed.}.
The ensuing mean mass is
$<\mu^1>/<\mu^0>=0.27~M_{\odot}$, assuming a standard spherical halo model.
When taking for the duration $T$ the values 
corrected for ``blending'', we get as average mass 0.34 $M_{\odot}$.
Although this value 
is marginally consistent with the result of the MACHO team,
it definitely favours a lower average MACHO mass.

For the fraction of the local
dark mass density detected
in the form of MACHOs, we find $f \sim 0.54$, which compares quite well
with the corresponding value ($f \sim 0.45$) calculated
by the MACHO group in a different way. 
However, the uncertainties on $f$ are large, due to the lack of
precise knowledge on the actual shape of the dark halo and its total
mass. 

\section{Formation of dark clusters}

A major problem concerns the formation of MACHOs, as well
as the nature of the remaining amount of dark matter in the galactic halo.
We feel it hard to conceive a formation mechanism which transforms with 100\%
efficiency hydrogen and helium gas into MACHOs. Therefore, we expect
that also cold clouds (mainly of $H_2$) should be present in the 
galactic halo. Recently, we have proposed
a scenario \cite{de,de1,de2,de3,de4}
in which dark clusters of MACHOs and cold molecular 
coulds naturally form in the halo at galactocentric distances
larger than 10--20 kpc, with the relative abundance possibly
depending on the distance.

The evolution of the primordial proto globular cluster clouds
(which make up the proto-galaxy)
is expected to be very different 
in the inner and outer parts of the Galaxy, depending on the decreasing 
ultraviolet flux (UV) from the centre
as the galactocentric distance $R$ increases.
In fact, in the outer halo no substantial $H_2$ depletion should 
take place, owing to the distance suppression of the UV flux.
Therefore, the clouds cool and fragment - the process stops when the 
fragment mass becomes $\sim 10^{-2} - 10^{-1}~M_{\odot}$.
In this way dark
clusters should form, which contain brown dwarfs  
and also cold $H_2$ self-gravitating cloud, 
along with some residual diffuse gas (the 
amount of diffuse gas inside a dark cluster has to be low, for otherwise it 
would have been observed in the radio band).

We have also considered several observational tests for our model 
\cite{de,di}.
In particular, a signature for the presence of 
molecular clouds in the galactic halo should be a $\gamma$-ray flux 
produced in the scattering of high-energy cosmic-ray protons on $H_2$.
As a matter of fact, an essential
information is the knowledge of the cosmic ray flux in the halo. Unfortunately,
this quantity is unknown and the only available 
information comes from theoretical considerations.
Nevertheless, we can make an estimate of the expected $\gamma$-ray flux
and the best chance to detect it is provided
by observations at high galactic latitude.
Accordingly, we find a $\gamma$-ray flux (for $E_{\gamma}>100$ MeV)
$\Phi_{\gamma}(90^0) \simeq ~\tilde f~(0.4 - 1.8) \times 10^{-5}$ 
photons cm$^{-2}$
s$^{-1}$  sr$^{-1}$ ($\tilde f$ stands for 
the fraction of halo dark matter in the form of gas),
if the cosmic rays are confined in the galactic halo, otherwise, if they
are confined in the local galaxy group \cite{kn:ber} 
$\Phi_{\gamma}(90^0) \simeq ~\tilde f~(0.6 - 3) \times 10^{-7}$ 
photons cm$^{-2}$
s$^{-1}$  sr$^{-1}$. 
These values should be compared with the measured
flux by the SAS-II satellite for the diffuse background of
$(0.7-2.3)\times 10^{-5}$ photons cm$^{-2}$ s$^{-1}$  sr$^{-1}$
or the corresponding flux found by EGRET of 
$\sim 1.1 \times 10^{-5}$ photons cm$^{-2}$ s$^{-1}$  sr$^{-1}$. Thus, 
there is at present no contradiction with observations.
Furthermore, an improvement of sensitivity for the next generation of 
$\gamma$-ray detectors will allow to clarify the origin
of this flux   
or yield more stringent limits on $\tilde f$.\\

\end{document}